\documentclass[showpacs,twocolumn,floatfix,amsmath,amssymb,superscriptaddress,prl,nopacs]{revtex4-1}
\usepackage{epsfig}
\usepackage{graphicx}
\usepackage{longtable}
\usepackage{color}

\usepackage{bm}

\newcommand{\bea}{\begin{eqnarray}}
\newcommand{\eea}{\end{eqnarray}}
\newcommand{\bt}{\textbf}

\newcommand{\phd}{\phantom{\dag}}
\newcommand{\ph}{\phantom{.}}
\newcommand{\up}{^{\phd}}
\newcommand{\noi}{\noindent}

\begin{document}

\title{Magnetic-Field-Induced Topological Reorganization of a P-wave Superconductor}
\author{Maria Teresa Mercaldo}
\affiliation{Dipartimento di Fisica ``E. R. Caianiello", Universit\`a di Salerno, IT-84084 Fisciano (SA), Italy}
\author{Mario Cuoco}
\affiliation{CNR-SPIN, IT-84084 Fisciano (SA), Italy}
\affiliation{Dipartimento di Fisica ``E. R. Caianiello", Universit\`a di Salerno, IT-84084 Fisciano (SA), Italy}
\author{Panagiotis Kotetes}
\affiliation{Center for Quantum Devices, Niels Bohr Institute, University of Copenhagen, 2100 Copenhagen, Denmark}
\affiliation{Institut f\"{u}r Theoretische Festk\"{o}rperphysik, Karlsruhe Institute of Technology, 76131 Karlsruhe, Germany}
\email{kotetes@nbi.ku.dk}
\begin{abstract}
In this work we illustrate the detrimental impact of the Cooper pair's spin-structure on the thermodynamic and topological pro\-per\-ties of a spin-triplet superconductor in an applied Zeeman field. We particularly focus on the paradigmatic one-dimensional case (Kitaev chain) for which we self-consistently retrieve the energetically preferred Cooper pair spin-state in terms of the corresponding spin $\bm{d}$-vector. The latter undergoes a substantial angular and amplitude reorganization upon the variation of the strength and the orientation of the field and results to a modification of the bulk topological phase diagram. Markedly, when addressing the open chain we find that the orientation of the ${\bm{d}}$-vector varies spatially near the boundary, affecting in this manner the appearance of Majorana fermions at the edge or even altering the properties of the bulk region. Our analysis reveals the limitations and breakdown of the bulk-boundary correspondence in interacting topological systems. 
\end{abstract}

\maketitle

{\textit{Introduction.}} Interacting quantum matter constitutes a realm of puzzling phenomena whose understan\-ding promises technolo\-gi\-cal breakthroughs. A striking exam\-ple is found in the so-called p-wave superconductors (PSCs) in which electrons form Cooper pairs in the symmetric spin-triplet and orbitally-antisymmetric configuration \cite{SigristUeda}. Substantial experimental investigation has provided evidence for spin-triplet pairing in diverse classes of materials~\cite{Maeno2012} 
(e.g. heavy fermions~\cite{Stewart1984,Saxena2000,Kyogaku1993}, non-centrosymmetrics~\cite{Bauer2004,Nishiyama2007}, organics~\cite{Lebed2000,Lee2001,Shinagawa2007}, oxides~\cite{Maeno1998}, topological superconductors~\cite{Matano2016}, etc.) and in a va\-rie\-ty of heterostructures consisting of spin-singlet SCs interfaced with magnetic systems~\cite{Buzdin,Efetov,Eschrig,Linder2015}. One tantalizing perspective of this search is fabricating devices for to\-po\-lo\-gi\-cal quantum computing based on Majorana fermions (MFs)~\cite{Kitaev2003,NayakReview,ReadGreen,KitaevChain,Ivanov,AliceaNatPhys,Milestones}. The latter correspond to neutral quasiparticle excitations which obey non-abelian exchange statistics enabling quantum operations by braiding them around one another~\cite{Ivanov,NayakReview,AliceaNatPhys}. They naturally emerge in one-dimensional (1d) spinless PSCs \cite{KitaevChain} which constitute the prototypical topological SCs (TSCs). Remarkably, every TSC can be mapped to a PSC in an appropriate limit \cite{Altland,KitaevClassi,Ryu}, a connection which motivated the proposal of artificial TSCs~\cite{FuKane2008,AliceaReview,BeenakkerReview,FlensbergReview,KotetesClassi} and the subsequent observation of MFs in hybrid superconducting devices~\cite{Mourik,Deng,Furdyna,Heiblum,Finck,Churchill,Yazdani,Franke,Pawlak,Albrecht}.

While the so far conducted analysis of PSCs has lead to tremendous progress and insight regarding topological systems in general, fundamental aspects of the system itself still remain unexplored. One of the main questions relates to the order parameter (OP) of the PSC, $\bm{d}$~\cite{SigristUeda}, which constitutes a vector in spin-space and its structure controls the topological behavior of the system. As a spin-vector, the OP is expected to exhibit a rich interplay with an externally applied magnetic field. Despite the immedia\-te experimental sig\-ni\-fi\-can\-ce, PSCs have been considered to be spinless in many cases, a simplification which nevertheless was proven to be an indispensable tool for topological purposes~\cite{KitaevChain}. On the other hand, existing insightful topological studies of spinful PSCs in external magnetic fields have either entirely \cite{Dumitrescu2013,Dumitrescu2014} or partially \cite{Hyart2014} neglected the energy stability of the various OP configurations, and thus the intrinsic complexity of the \textit{spatially} resolved self-consistent solutions. In particular, a realistic aspect that has been completely overlooked is the role of boun\-dary effects either locally by directly affec\-ting the occurrence of MFs or globally via their feedback on the bulk or/and topological properties.   

In this Letter, we show that the Cooper pair spin-configuration of a 1d PSC with an easy spin-plane and chiral symmetry \cite{Dumitrescu2014,Tewari and Sau, ChiralTanaka,SatoChiral,NOZeemanPK} experiences an intricate rearrangement in response to an applied Zeeman field ($\bm{h}$), with dramatic consequences on the to\-po\-lo\-gi\-cal properties. The non-self-consistent topological phase diagram, con\-si\-sting of phases with ${\cal N}=1,2$ MFs per edge, becomes signi\-fi\-cant\-ly modified when one naturally {\color{black}{allows the $\bm{d}$-vector to reorganize in such a way so}} to minimize the free energy. We reveal that this internal degree of freedom opens the path to new topological phases with \textit{enhanced} (${\cal N}=3,4$) number of MFs per edge. More strikingly, for an open chain we encounter the bulk-boundary correspondence breakdown, since the number of MFs per edge does not coincide with the bulk topological invariant ${\cal N}$. This stems from the inhomogeneous profile of $\bm{d}$ near the boundary and the accompanying induced magnetization $\bm{M}\propto i\bm{d}\times\bm{d}^*$~\cite{Hyart2014}. In fact, the system be\-ne\-fits ener\-getically from substantially reorganizing the OP near the boundary to such an extent, that it even affects and reconstructs the OP in the bulk. Beyond the easy plane picture, i.e. by assuming $d_z\neq0$, we find that boundary effects can spontaneously violate chiral symmetry. Thus, phases with an even number of MFs per edge become elusive in a rea\-li\-stic situation, further invigorating the unreliability of the bulk-boundary correspondence predictions in \textit{correlated} topological matter. 
\begin{figure}[t]
\centering
\includegraphics[width=0.8\columnwidth]{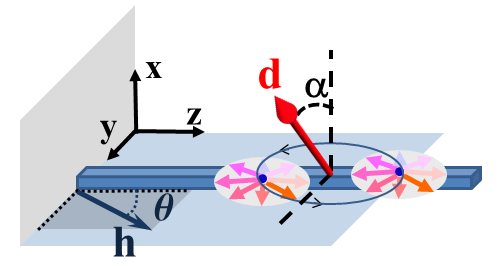}
\caption{1d spin-triplet p-wave superconductor in an applied Zeeman field ($\bm{h}$). We consider an easy $xy$ spin-plane ($\alpha$ angle) for the spin $\bm{d}$-vector order parameter, while $\bm{h}$ is confined to the $yz$ plane ($\theta$ angle). The additional arrows illustrate the orientation of the Cooper pair spin-moment for the given $\bm{d}$-vector.}
\label{fig:CartoonSystem}
\end{figure}

{\textit{Model Hamiltonian.}} The Bogoliubov - de Gennes (BdG) Hamiltonian for the con\-si\-de\-red bulk 1d PSC reads
\bea
\widehat{{\cal H}}_k=\varepsilon_k\tau_z-\bm{h}\cdot\bm{\sigma}+\tau_+\ph\bm{d}_k\cdot\bm{\sigma}+\tau_-\ph\bm{d}_k^*\cdot\bm{\sigma}\,,\quad\label{eq:Hamiltonian}
\eea

\noi where ${\cal H}=\frac{1}{2}\sum_k\Psi_k^{\dag}\widehat{\cal H}_k\Psi_k$. We used the Nambu spinor $\Psi_k^{\dag}=(\psi_{k\uparrow}^{\dag}\,,\psi_{k\downarrow}^{\dag}\,,\psi_{-k\downarrow}\up\,,-\psi_{-k\uparrow}\up)$ and the $\bm{\sigma}$ ($\bm{\tau}$) spin (particle-hole) Pauli matrices. We assume the electron dispersion $\varepsilon_k=-2t\cos(ka)-2t'\cos(2ka)-2t''\cos(3ka)-\mu$ \cite{hoppings} and $t=1$, with $t^{\nu}$ denoting hopping to the $\nu$-th nearest neighbor with lattice constant $a=1$ and $\mu$ is the che\-mi\-cal potential. In addition, we introduced the Zeeman field $\bm{h}$ and the odd-parity OP $\bm{d}_k=2\bm{d}\sin k$, with $\bm{d}$ the earlier mentioned complex vector defining the spin-orientation of the OP. One can also define a matrix OP in spin-space $\{\uparrow,\downarrow\}$, $\widehat{\Delta}=\bm{d}\cdot\bm{\sigma}\sigma_y$: $\Delta_{\uparrow\uparrow,\downarrow\downarrow}=d_y\pm id_x$ and $\Delta_{\uparrow\downarrow}=-id_z$. 

The considered PSC (Fig.\ref{fig:CartoonSystem}) {\color{black}{has}} the main axis along the $z$ direction and a square $xy$ cross-section, and is assumed invariant under the ${\rm D_{4h}}$ point group. While we neglect spin-orbit interaction, we still allow the spin-vectors to transform under ${\rm D_{4h}}$. We further consider an effective separable four-fermion interaction in the PSC channel with anisotropic potentials $V_x=V_y\equiv V$ and $V_z$ for the spin- $xy$ plane and $z$ axis, respectively \cite{SupMat}. We determine the ground-state (both for closed and open chains) based on an iterative self-consistent scheme of computation and on the minimization of the total free energy \cite{SupMat,Cuoco1,Cuoco2}. Our investigation was performed for a representative amplitude $V=2$ and for different chain sizes $L$. Modifying the latter leaves the results qualitatively unchanged and here we focus on $L=300$.
\begin{figure*}[t]
\centering
\includegraphics[width=1.0\textwidth]{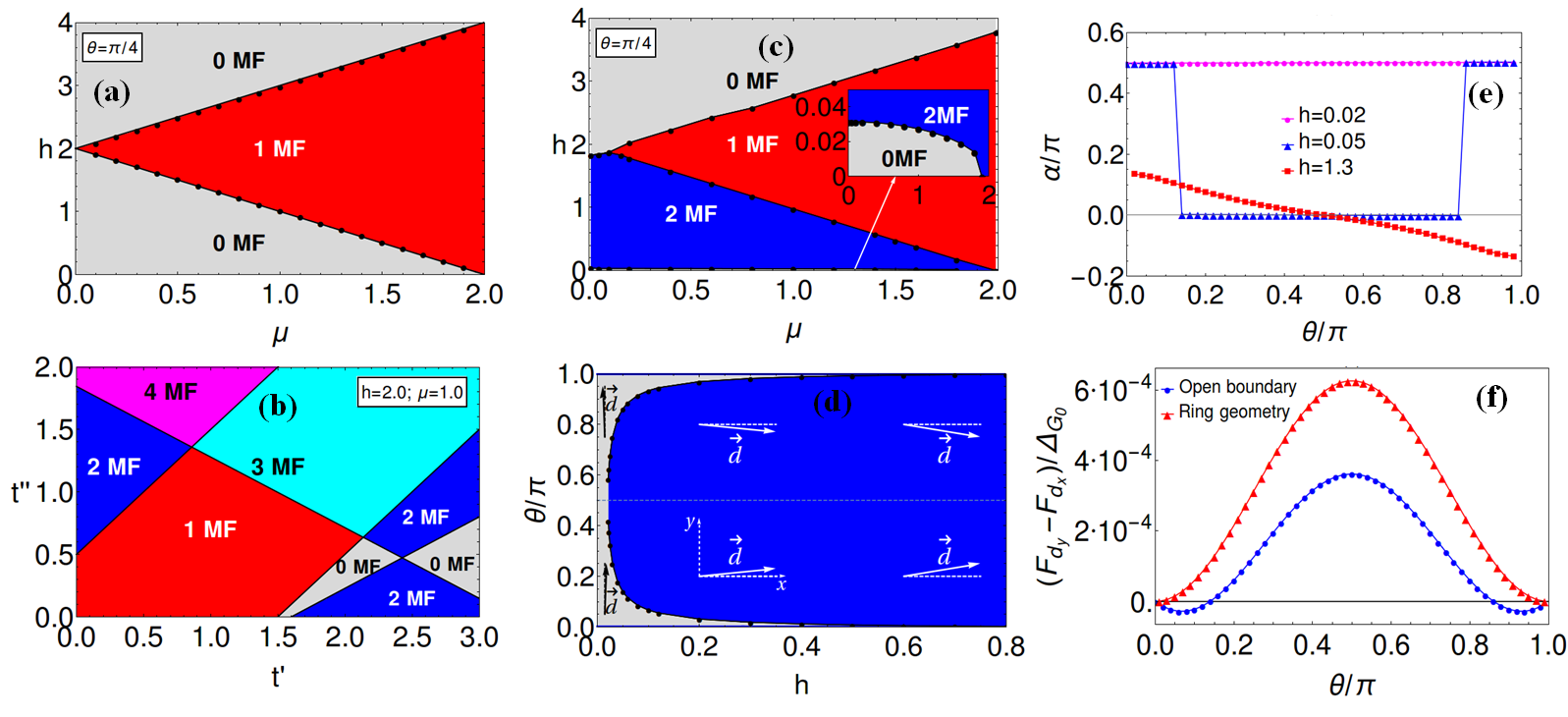}
\caption{a) Non-self-consistent phase diagram of the 1d PSC as a function of $h$ and $\mu$ at a given field orientation $\theta=\pi/4$, while $\bm{d}=(0,d_y,0)$. (b) Topological phase diagram in terms of the 2nd and 3rd nearest neighbour hopping for $h=2$, $\theta=\pi/4$ and $\bm{d}=(d_x,d_y,0)$. The addition of $d_x$ allows for \textit{new} phases with ${\cal N}=3,4$. (c) Self-consistent topological phase diagram for an open chain for $\theta=\pi/4$. In contrast to (a), the ${\cal N}=2$ phase becomes accessible. (d) Ground-state topological phase diagram in terms of $\{\theta,h\}$ for an open chain, with $\mu=0.6$. The arrows schematically depict the $\bm{d}$-vector orientation. Gray and black areas correspond to phases with 0 and 2 MFs per edge. (e) Dependence of the polar angle $\alpha$, with $\bm{d}=d(i\cos\alpha,\sin\alpha,0)$, on the field orientation for different strengths of $h$ and fixed $\mu=0.6$. (f) Difference between the total free energies $F_{d_y}$ and $F_{d_x}$, corresponding to the self-consistent $\bm{d}$-vector solutions with dominant $d_y$ and $d_x$ components, for an open boundary and closed ring geometry as a function of the field orientation $\theta$. Starkingly, the boundary reorganization of $\bm{d}$ drives a bulk transition (f) with the \textit{fully}-self-consistent phase diagrams of (c,d). Here $\mu=0.6$, $h=0.05$ and $\Delta_{G_0}$ is the gap for $h=0$.}
\label{fig:PDESP}
\end{figure*}

{\textit{Bulk non-self-consistent topological phase diagram for $d_z=0$.}} We start with the \textit{bulk} case (infinite system or ring geometry) and $d_z=0$. 
At zero field we can arbitrarily choose the orientation of $\bm{d}$ in the $xy$ spin-plane. The Hamiltonian decomposes into two spin blocks, each of which contributes with 1MF per edge if $|t+t''|>|t'+\mu/2|$, as long as bulk-boundary correspondence applies. If we keep the orientation of the $\bm{d}$-vector frozen we find that the application of a parallel Zeeman field renders the PSC topologically trivial. In stark contrast, for a perpendicular field the topological regime occurs when $|t+t''|>|t'+(\mu\pm h_{\perp})/2|$ ($h_{\perp}=|\bm{h}_{\perp}|$) and we retrieve TSC phases with topological invariant ${\cal N}=1,2$. We thus conclude that as long as the field {\color{black}{has}} a perpendicular component to the $\bm{d}$-vector, MFs become accessible. For an arbitrary orientation of the field in the $yz$ plane the Hamiltonian resides in the BDI symmetry class with chiral, time-reversal and charge-conjugation symmetries \cite{Altland,KitaevClassi,Ryu} with corre\-spon\-ding ope\-ra\-tors: $\Pi=\tau_x\sigma_x$, $\Theta=\tau_z\sigma_z{\cal K}$ and $\Xi=\tau_y\sigma_y{\cal K}$. ${\cal K}$ defines complex conjugation. The topological phase boun\-daries are determined by the spectrum gap closings at the inversion symmetric $k_{{\cal I}}=0,\pi$ points, occuring for $|\varepsilon_{k_{{\cal I}}}|=h$ (Fig.~\ref{fig:PDESP}(a)). Interestingly, while the ${\cal N}=1,2$ phases remain in principle accessible, only the ${\cal N}=1$ appears.

{\textit{Bulk self-consistent topological phase diagram for $d_z=0$.}} While the results presented above are certainly insightful, they cannot capture the actual behavior of a PSC in the presence of the field. The loophole relies on the assumption that the spin-direction of the $\bm{d}$-vector is frozen. The latter is against the self-consistent nature of the OP and does not account for its freedom to readjust in order to minimize the free ener\-gy. The first general conclusion obtained from the self-consistent analysis is that the $\bm{d}$-vector always prefers to be \textit{perpendicular} to the field. Therefore, in the spin-isotropic case $V_x=V_y=V_z$ the system would always reorganize so to ensure $\bm{d}\perp\bm{h}$ and lead to TSC phases with ${\cal N}=1,2$. However, in the presence of an easy $xy$ spin-plane the situation becomes more intricate since there is an additional energy cost to be paid for achieving $\bm{d}\perp\bm{h}$, as a result of the spin-anisotropic interaction. Therefore, the configuration $\bm{d}\perp\bm{h}$ does not ne\-ces\-sa\-ri\-ly correspond to the ground state of the system and the topological phase diagram becomes much richer. 

For the purpose of highlighting the a\-ri\-sing be\-ha\-vior we focus on the extreme anisotropic limit, i.e. $V_z=d_z=0$. At zero field the $\bm{d}$-vector can freely rotate in the $xy$ plane. When the field is turned on, the $\bm{d}$ vector will try to maximize the term $|\bm{d}\times\bm{h}|$. For $V_z=0$, implying $\bm{d}=(d_x,d_y,0)\equiv d(i\cos\alpha,\sin\alpha,0)$ and $\bm{h}=(0,h_y,h_z)$, we find that the $d_x$ component is always present since it is perpendicular to the field. The $d_y$ component vani\-shes if $h_z=0$ ($\theta=\pm\pi/2$) and is de\-ge\-ne\-ra\-te with $d_x$ for $h_y=0$ ($\theta=0,\pi$). For the remaining values of $\theta$, $d_y$ is non-zero and becomes sizeable when $h_z\gg h_y$. The bulk analysis further shows that the field-induced reorganization of the $\bm{d}$-vector is strongly dependent on the amplitude of the field. In the weak field regime the $\bm{d}$-vector's angle, $\alpha$, exhibits a pronounced dependence on the field angle $\theta$, and it becomes pinned to $\alpha=\pi/4$ above a thre\-shold field correspon\-ding to full spin-polarization. 

Remar\-kably, the simul\-ta\-ne\-ous pre\-sen\-ce of the two components of the $\bm{d}$-vector, apart from modifying the boundaries of the preexisting ${\cal N}=1,2$ TSC phases, can also \textit{radically} mo\-di\-fy the phase diagram by introdu\-cing new phases with ${\cal N}>2$ due to next-nearest neighbors hopping. The origin of such an effect can be attributed to the modified gap closing conditions of the energy spectrum, $\varepsilon_k^2=h^2+4d^2\cos(2\alpha)\sin^2k$ and $\sin k\left(\varepsilon_k-h_z\tan\alpha\right)=0$, which provide the topological phase boundaries. The energy spectrum not only exhibits gap closings at $k_{{\cal I}}$, but also for the $\pm k_*$ points determined by $\varepsilon_{k_*}=h_z\tan\alpha$. Each gap closing for a $\pm k_*$ pair changes ${\cal N}$ by 2, allowing phases with ${\cal N}>2$ and additional pairs of MFs per edge protected by chiral symmetry \cite{Altland,KitaevClassi,Ryu,Dumitrescu2014,Tewari and Sau,ChiralTanaka,SatoChiral,NOZeemanPK}. 

For the given $\varepsilon_k$ and $\bm{d}_k$, we find ${\cal N}=3,4$ (see Fig.~\ref{fig:PDESP}(b)), which as long as bulk-boundary correspondence is intact implies 3,4MFs per edge. In fact, the phase with 3(4) MFs per edge can be accessed from a 1(2) MF phase and can be captured only within the self-consistent approach which imposes the two components of the $\bm{d}$-vector, {\color{black}{as explicitly shown in Ref. \cite{SupMat}}}. Here phases with ${\cal N}=3,4$ occur in a small region of the parameter space, implying that such a scenario may not be easily accessible in the particular materials. Nonetheless, such a MF-induction-mechanism without altering the symmetry class is quite generic and can be pursued in alternative topological systems.

{\textit{Boundary effects on the topological phase diagram for $d_z=0$.}} While one would expect that, according to the bulk-boundary correspondence, the self-consistently performed \textit{bulk} analysis of the topological invariant ${\cal N}$ would provide the number of MFs appearing per edge, such a connection becomes invalid here. As a matter of fact, cases of bulk-boundary correspondence breakdown become ma\-ni\-fest when we determine the ground-state by performing a self-consistent ana\-ly\-sis using open, instead of periodic, boundary conditions. For weak applied fields we obtain a first order transition resulting to a rotation of the $\bm{d}$ vector, due to an emergent competition between boundary and bulk contributions to the free energy. Bulk-boundary correspondence becomes systematically inapplicable due to a reorientation of the $\bm{d}$-vector all over the system, which allows to the $d_y$ component to dominate over the $d_x$ for $\theta\sim 0\,,\pi$. This result corrects the bulk self-consistent approach and renders, in any realistic situation, part of the previously considered to\-po\-lo\-gi\-cal regime as trivial. The resulting phase diagram as a function of the amplitude of the applied field and the electron density is shown in Fig.~\ref{fig:PDESP}(c), for a fixed representative orientation of the field. The increase of the field drives a reo\-rien\-ta\-tion of the $\bm{d}$-vector via a first order $d_y\rightarrow d_x$ transition (Figs.~\ref{fig:PDESP}(d),(e)) concerning the dominant $\bm{d}$-vector component. Such transition depends on the strength of the field and vanishes above a thre\-shold which for the pa\-ra\-me\-ters assumed is $h\sim 0.8$. The region of the phase diagram with $|d_x|\gg |d_y|$ is topologically non-trivial and supports 2MFs per edge. 

The detailed investigation of the free energy evolution for the two solutions, reveals that the stability of the configuration with a dominant $d_y$ relies on the free energy lowering via the OP edge reconstruction. As we mentioned above, for a ring geometry there exists a degeneracy between $d_{x,y}$ for $\theta=0$, while for $\theta\neq0$ the $d_y$ component is disfavored. Hence, as depicted in Fig.~\ref{fig:PDESP}(f), the energy exhibits a monotonic behavior as $\theta$ varies in $[0,\frac\pi 2]$. In contrast, we find that in the presence of the boun\-dary the free energy develops minima with $|d_y|\gg |d_x|$ for $\theta\sim0\,,\pi$. We have also investigated the spatial profile of the $\bm{d}$-vector and verified that there is a substantial increase of the $d_y$ component near the edge. Such a boundary OP reorganization appears capable of modifying the properties of the entire SC and conclude with a topological transition without any gap closing. The above mentioned transition can be alternatively viewed as a result of the interplay between the magnetization near the boundary and the spin-triplet OP. The emergent spin-activity of the boundary can relate to $\bm{d}$-vector reconstruction phenomena encountered in ferromagnet-PSC hybrids, which are understood on the basis of spin-filtered Andreev reflection \cite{Gentile2013}.
\begin{figure}[t]
\begin{center}
\includegraphics[width=1.0\columnwidth]{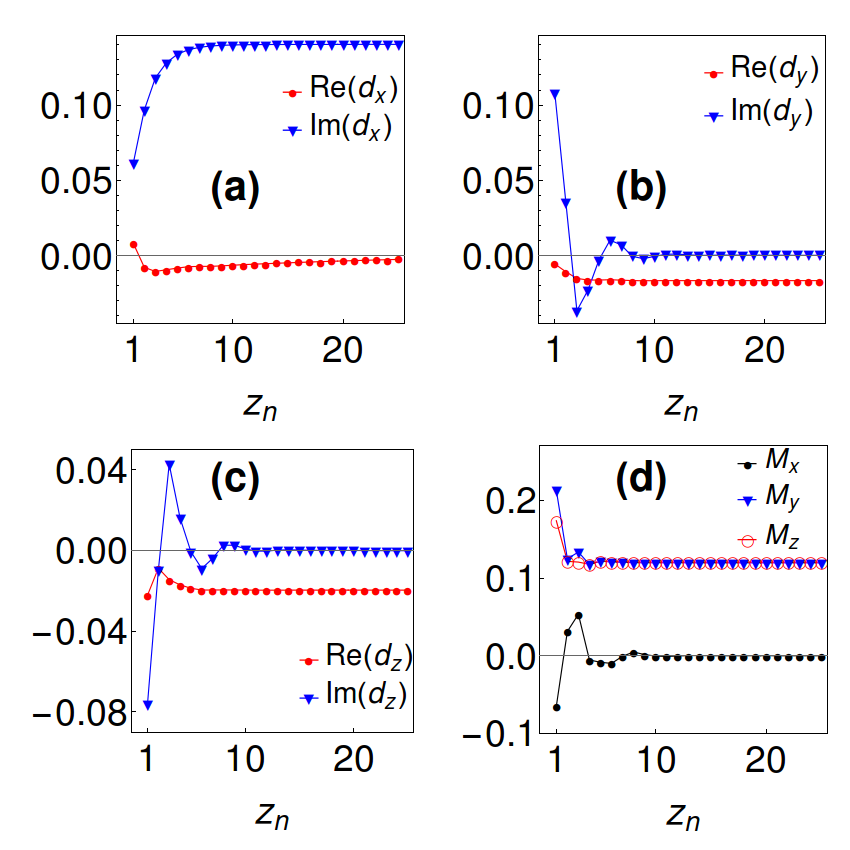}
\end{center}
\caption{Self-consistent edge profile of the complex (a) $d_x$, (b) $d_y$, (c) $d_z$ spin-triplet order parameter components and (d) the magnetization ($\bm{M}$), for $\mu=0.6$, $V_z=0.5$ and a given amplitude ($h=1$) and orientation of the applied field ($\theta=\pi/4$). The $M_x$ appearing near the boundary spontaneously violates the bulk chiral symmetry and drives the bulk-boundary correspondence breakdown.}
\label{fig:3}
\end{figure}

{\textit{Self-consistent topological phase diagram for $d_z\neq0$.}} In the case of an easy axis, $V_z\gg V_{x,y}$, the direction of the $\bm{d}$-vector is essentially frozen and our non-self-consistent analysis applies: when the field is parallel to the $z$ axis the system is a topologically-trivial PSC, while when a field-component perpendicular to the $\bm{d}$-vector appears, MFs become in principle accessible. Only the case $V_z\ll V$ remains to be explored. In the latter, the $d_z$ OP is expected to be small and as long as the field is constrained within the $yz$ plane, the ap\-pea\-ran\-ce of $d_z$ is mainly controlled by the following magnetic contribution $ih_y(d_z^*d_x-d_zd_x^*)$ to the free energy \cite{Hyart2014}, which sources a non-zero bulk magnetization $M_y$. Therefore, $d_z\propto ih_yd_x\propto i\sin\theta d_x$. The presence of a bulk $d_z$ component with the given phase locking respects the preexesting chiral symmetry of the system, since the corresponding Hamiltonian term reads $2d_z\sin k\tau_x\sigma_z$ ($d_z\in\mathbb{R}$). Therefore the TSC phases ${\cal N}=1,2$ are still accessible.

While the above picture is indeed confirmed by our self-consistent calculations  when imposing perio\-dic boundary conditions, severe discrepancies emerge in the presence of boundaries related to the spontaneous appearance of an $M_x\propto i(d_y^*d_z-d_yd_z^*)$ magnetization component. The latter becomes non-zero \textit{solely} near the boundary since only there the $\bm{d}$-vector becomes complex, while in the bulk $d_y$ and $d_z$ are both real implying $M_x=0$. The spatial profiles of $\bm{d}$ and $\bm{M}$ for an open chain are shown in Fig.~\ref{fig:3}. The complex character of the $d_{y,z}$ components near the boundary leads to the respective spontaneous violation of chiral symmetry. Thus we arrive to another {\color{black}{case}} of bulk-boundary correspondence violation, since phases with ${\cal N}=2$ which would imply 2MFs per edge, now become unobservable as each MF pair hybridizes into finite energy bound states. 

{\textit{Conclusions.}} As we showed above, spin-triplet superconductors in a Zeeman field actually exhibit to\-po\-lo\-gi\-cal sce\-na\-rios which are completely unexpected within a non-self-consistent framework and crucially depend on the order parameter's spin-structure and spatial-profile. Our self-consistent analysis revealed that such materials exhibit topological shielding, i.e. the increase of the topological invariant in order to minimize the system's free energy. Moreover, the substantial boundary reconstruction of the order parameter can also drastically affect the appearance of Majorana fermions. The latter either occurs due to the spontaneous violation of a particular bulk symmetry near the boundary or the ra\-di\-cal reconstruction of the bulk ground state when the boundary contribution to the free energy becomes sig\-ni\-fi\-cant. These interaction-driven manifestations of bulk-boundary correspondence breakdown open perspectives for driving a topological phase transition without any bulk spectrum gap closing but rather via solely controlling the physical properties at the edge.

\end{document}